\documentclass[11pt]{article}

\usepackage{amsmath}
\usepackage{amssymb}
\usepackage{graphicx}
\usepackage{hyperref} 
\usepackage{geometry} 
\usepackage{doi}      
\usepackage{adjustbox}
\usepackage{pdflscape}
\usepackage{multirow}

\usepackage[russian, english]{babel}

\usepackage{lscape}

\title{Multimodal Stock Price Prediction:\\A Case Study of the Russian Securities Market}
\author{
  Kasymkhan Khubiyev \\
  Sirius University of Science and Technology, Sirius, Russia \\
  \texttt{kasymkhankhubievnis@gmail.com}
\and
  Mikhail Semenov\\
  Sirius University of Science and Technology, Sirius, Russia \\
  \texttt{semenov.me@talantiuspeh.ru}
}
\date{\today}

\begin{document}

\maketitle

\footnotetext{Preprint for the \textit{Program Systems: Theory and Applications} journal
}

\begin{abstract}
Classical asset price forecasting methods primarily rely on numerical data, such as price time series, trading volumes, limit order book data, and technical analysis indicators. However, the news flow plays a significant role in price formation, making the development of multimodal approaches that combine textual and numerical data for improved prediction accuracy highly relevant.

This paper addresses the problem of forecasting financial asset prices using the multimodal approach that combines candlestick time series and textual news flow data. A unique dataset was collected for the study, which includes time series for 176 Russian stocks traded on the Moscow Exchange and $79,555$ financial news articles in Russian.

For processing textual data, pre-trained models RuBERT and Vikhr-Qwen2.5-0.5b-Instruct (a large language model) were used, while time series and vectorized text data were processed using an LSTM recurrent neural network. The experiments compared models based on a single modality (time series only) and two modalities, as well as various methods for aggregating text vector representations.

Prediction quality was estimated using two key metrics: Accuracy (direction of price movement prediction: up or down) and Mean Absolute Percentage Error (MAPE), which measures the deviation of the predicted price from the true price. The experiments showed that incorporating textual modality reduced the MAPE value by 55\%.

The resulting multimodal dataset holds value for the further adaptation of language models in the financial sector. Future research directions include optimizing textual modality parameters, such as the time window, sentiment, and chronological order of news messages.

\end{abstract}

\section{Introduction}
Building a price forecast for an asset is a crucial task for financial market participants, as it enables strategic planning, optimal investment portfolio management, and risk assessment. Numerous attempts have been made to apply machine learning methods to construct such forecasts~\cite{jaggitext(2021), mishevevaluation(2020), hostock(2021)}. With the growing popularity of deep learning models, researchers have shifted their focus toward the application of neural networks. At the same time, the problem of accurately accounting for the news flow as a key factor influencing market behavior is being reconsidered with the rapid development of generative artificial intelligence models and large language models (LLMs) such as ChatGPT, FinGPT, GigaChat, LLama, and others. In financial economics, LLMs are still rarely used, and their full potential remains untapped.

Researchers are exploring the use of natural language processing models to enhance the accuracy of asset price forecasts and investment portfolio management strategies.  

The study~\cite{Bledar(2022)} describes the use of sentiment analysis of news as an additional parameter. The authors employed the FinBert model, trained on financial data, to assess the sentiment of news articles as positive, negative, or neutral. The study utilized time series data from candlestick charts of the U.\,S. stock market index, Standard \& Poor’s 500 (S\&P 500). A machine learning model — random forest —was used for price prediction.  
The study concluded that incorporating sentiment analysis of news flow improves prediction accuracy.

In the study~\cite{Yu(2023)}, the authors aimed to develop a multimodal artificial intelligence model capable of providing well-founded and accurate forecasts for time series data. They implemented a model that generates predictions of an asset’s monthly or weekly returns, accompanied by a textual explanation from a language model based on the user's input query.  

The study~\cite{Zhang(2023)} proposed an approach for fine-tuning instructions to interpret numerical values and contextualize financial data.
Kulikova et al.~\cite{Kulkova(2023)} examined the effect of classifying news into thematic groups. The authors demonstrated that, in most cases, it is advisable to use a single thematic group of news for the deep learning models considered (Temporal Convolutional Network, D-Linear, Transformer, and Temporal Fusion Transformer). They also determined the probabilities of forecast improvement for the 20 thematic groups analyzed.  

In all the aforementioned studies, the models were implemented using a multimodal approach for the U.\,S. stock market, with English as the modality language. Notably, the news flow was not integrated directly into the predictor’s input vector but rather through a preprocessing block in the form of an additional parameter, such as sentiment analysis, news frequency related to the asset, or news classification, etc.

The objective of the current study is to demonstrate the advantages of a new multimodal method over predictions based solely on numerical data and to present a Russian-language financial news dataset.  

To achieve this objective, we formulated the following key tasks:  
\begin{enumerate}
\item Construct a multimodal dataset consisting of time series data and news articles.  
\item Develop a predictive model capable of utilizing one or two modalities.  
\item Train the predictive model and analyze the values of accuracy functions and metrics, specifically Accuracy and MAPE.
\end{enumerate}

In this study, we propose a new multimodal approach for integrating news flow into time series numerical data. The text of the news articles is converted into a vector representation and fed into the model alongside the time series vector.  

Our hypothesis is that the multimodal approach will enable predictive models to extract semantic information from the text, thereby improving the accuracy of asset price forecasts.

\section{Data Collection and Structuring}

Multimodality implies the use of more than one data modality, which affects both the data structure and the logic of predictive model development. We utilize two types of modalities: (a) numerical — time series of stock prices, and (b) textual — news streams. To train the predictive model and analyze its performance, we collected an original dataset.  

The time series, represented as candlestick data with open, close, high, and low prices, were obtained through the Algopack API of the Moscow Exchange (MOEX). For the numerical experiment, we selected stock time series data spanning from July 7, 2022, to August 30, 2024, covering 176 companies. During this period, the Russian stock market experienced phases of rapid growth and decline, with the IMOEX index rising from 2,213.81 to 2,650.32 points ($+19,72\%$).

We collected $79,555$ news articles from various sources, including the online publication "RBC" (1,823 articles), "BCS Express" (11,331), and "BCS Technical Analysis" (9,670), the investment company website "Finam" (20,647), the trader community website "SmartLab.ru" (30,857), as well as the Telegram channel "RDV" (5,227).  

Several factors justify the selection of these sources. First, they provide news coverage for the required time period. Second, the institutional differences between sources, along with variations in writing style and levels of expertise, contribute to a more objective representation of events related to the analyzed time series.

News messages were tokenized using two models: RuBERT~\cite{Kuratov(2019)} and Vikhr-Qwen2.5-0.5b-Instruct~\cite{nikolichvikhr(2024)} (further as Qwen). In the context of tokenized text, a word refers to a token -- an element of the vector space represented as an index in the tokenizer’s vocabulary.  
Descriptive statistics of the dataset (in tokens), including mean, standard deviation, minimum, maximum word count, and quartiles, are presented in Tables~\ref{table:article_token_statistic} and~\ref{table:article_token_statistic_qwen}. It is important to note that tokenization can increase the word count in a text, for example, by splitting words into smaller components.

\begin{table}[ht]
\centering
\begin{tabular}{|l|r|r|r|r|r|r|r|}
\hline
\multicolumn{1}{|c|}{Source} &
  \multicolumn{1}{c|}{Mean} &
  \multicolumn{1}{c|}{Std} &
  \multicolumn{1}{c|}{Min} &
  \multicolumn{1}{c|}{Max} &
  \multicolumn{1}{c|}{Q25} &
  \multicolumn{1}{c|}{Q50} &
  \multicolumn{1}{c|}{Q75} \\ \hline
RDV           & 134 & 88  & 8  & 512 & 65  & 123 & 187 \\ \hline
Finam         & 221 & 135 & 18 & 512 & 116 & 178 & 284 \\ \hline
BCS Express  & 20  & 10  & 4  & 82  & 13  & 17  & 26  \\ \hline
BCS Technical Analysis & 502 & 37  & 29 & 512 & 512 & 512 & 512 \\ \hline
RBC           & 43  & 7   & 16 & 75  & 39  & 44  & 48  \\ \hline
SmartLab      & 21  & 8   & 5  & 82  & 15  & 19  & 25  \\ \hline
\end{tabular}
    \caption{Statistical features of the dataset after tokenization, RuBert. 
    }\label{table:article_token_statistic}
\end{table}

\begin{table}[h!]
\centering
\begin{tabular}{|l|r|r|r|r|r|r|r|}
\hline
Source &
  \multicolumn{1}{c|}{Mean} &
  \multicolumn{1}{c|}{Std} &
  \multicolumn{1}{c|}{Min} &
  \multicolumn{1}{c|}{Max} &
  \multicolumn{1}{c|}{Q25} &
  \multicolumn{1}{c|}{Q50} &
  \multicolumn{1}{c|}{Q75} \\ \hline
RDV          & 215  & 157 & 3  & 1324 & 92   & 187  & 304  \\ \hline
Finam         & 453  & 405 & 35 & 5732 & 211  & 319  & 501  \\ \hline
BCS Express  & 36   & 19  & 5  & 163  & 23   & 32   & 47   \\ \hline
BCS Technical Analysis & 1493 & 310 & 40 & 2221 & 1448 & 1545 & 1665 \\ \hline
RBC           & 75   & 12  & 28 & 105  & 68   & 77   & 83   \\ \hline
SmartLab      & 33   & 12  & 7  & 120  & 25   & 31   & 39   \\ \hline
\end{tabular}
    \caption{Statistical features of the dataset after tokenization, Qwen.
}\label{table:article_token_statistic_qwen}
\end{table}

\selectlanguage{russian}
Table~\ref{table:tokenized_sentence_example} provides examples of how a phrase changes after tokenization. For instance, the word <<открывает>> is split into three subcomponents: <<от>>, <<\#\#к>>, and <<\#\#рывает>>, where the <<\#\#>> prefix indicates that the token is a continuation of the previous token.
\selectlanguage{english}

\begin{table}[ht]
\selectlanguage{russian}
    \centering
    \begin{adjustbox}{width=\textwidth}
        \begin{tabular}{|l|l|}
            \hline
            Original text & Tokenized text\\\hline
            Доллар снова ниже 69 рублей & До \#\#лла \#\#р снова ниже 69 рублей\\\hline
            Москвич банкрот? & Москви \#\#ч банк \#\#рот ?\\\hline
            НПО Наука Отчет РСБУ & Н \#\#П, \#\#О Наука От \#\#чет Р \#\#С \#\#Б \#\#У\\\hline
            T-банк это желтый банк & T - банк это же \#\#лт \#\#ый банк \\\hline
        \end{tabular}
    \end{adjustbox}
    \selectlanguage{english}
      \caption{Original and tokenized texts examples}\label{table:tokenized_sentence_example}
\end{table}

\textbf{News articles characteristics} On the "BCS Technical Analysis" platform, news articles tend to be lengthy, which imposes limitations on tokenizers. Specifically, as shown in Tables~\ref{table:article_token_statistic} and~\ref{table:article_token_statistic_qwen}, the RuBERT model truncates the tokenized vector for longer texts. Additionally, the average length of tokenized text using the Qwen model exceeds that of RuBERT, indicating that Qwen has a broader vocabulary and a stronger text decomposition capability.

Furthermore, we collected data on 176 companies, forming a dataset consisting of tuples in the format:
\begin{center}
«ticker - company name - company activity description».
\end{center}
Such data are essential in our case for: (a) extracting keywords from company descriptions, and (b) improving the language model’s ability to link events described in news articles to specific companies and assess the impact of news on price dynamics.  

The dataset of news articles includes the following parameters: publication date, source, title, article body, and tags (keywords). For sources such as "RDV" and "SmartLab", article titles are absent, and the corresponding fields are filled with a label: \textit{no title}.
In our case, tags may include the full or abbreviated company name along with the corresponding ticker, the name of the market sector, and similar information. Tags in news articles were assigned by the article authors.  

For the "RDV" source, tags were marked by authors in the form of hashtags 
\selectlanguage{russian}(e.\,g., \#цифры, \#аналитика). \selectlanguage{english}In "BCS Express" and "BCS Technical Analysis", tags were specified in dedicated fields at the beginning or end of the news article (e.\,g., PhoseAgro, Russian market) and were extracted from the \textit{HTML} code of the page using the corresponding \textit{HTML} tags. When tags were absent ("RBC", "SmartLab"), the parameter in the dataset remained empty.  

Table~\ref{table:article_tags} provides examples of news articles (headline fragments) along with their assigned tags.

\begin{table}[]
\selectlanguage{russian}
    \centering
    \begin{adjustbox}{width=\textwidth}
        \begin{tabular}{|l|l|l|}
            \hline
            Source & Article fragment (heading) & Tags\\\hline
            RDV & Сегежа (SGZH): таргет 16.2 руб., апсайд +102... & SGZH \\\hline
            RDV & Артген биотех (ABIO) завершил доклинические... & аналитика, ABIO\\\hline
            Finam & Индекс МосБиржи восстанавливает позиции и приб... & ФосАгро, ВСМПО-АВСМ, CNYRUB\\\hline
            Finam & <<Ашинский метзавод>> назвал АО "Урал-ВК"\, своим ... & АшинскийМЗ\\\hline
            BCS Express & <<Восходящее окно>>: в каких бумагах замечен это... & Селигдар SELG, ЕвроТранс EUTR\\\hline
            BCS Express & <<Сила Сибири>> выйдет на максимальную мощность... & Газпром GAZP\\\hline
            BCS Technical Analysis & Мечел. Что ждать от бумаг на следующей неделе & Мечел\\\hline
            BCS Technical Analysis & На предыдущей торговой сессии акции Норникеля ... & ГМК Норникель\\\hline
        \end{tabular}
    \end{adjustbox}
\selectlanguage{english}
     \caption{Examples of news articles (header snippet) and assigned tags}\label{table:article_tags}
\end{table}

\section{Methodology}
To validate our hypothesis regarding the advantages of the multimodal approach, we have planned a series of experiments.  

The first series of experiments focused on predicting prices using only numerical time series of candlestick characteristics (close, open, high, and low prices). The quality metrics obtained from this experiment serve as baseline values against which improvements in price prediction accuracy using the proposed multimodal approach will be evaluated.  

The second series of experiments aims to generate predictions and compute accuracy metrics (Accuracy, MAPE) using the multimodal approach while exploring different aggregation methods (Sum, Mean) for the vectorized news stream.

\subsection{The Single-Modality Approach}

We first conducted a series of experiments on asset price prediction using only time series data. For this, we applied classical machine learning models to the daily price values (close, open, high, low), including linear regression (LinReg), \textit{k}-nearest neighbors (KNN), decision tree (DT), random forest (RF), and the boosting algorithm XGBoost (XGB). Among deep learning models, we utilized a long short-term memory recurrent neural network (LSTM).

Conceptually, the experiment consists of two tasks: (a) predicting the price movement direction (increase or decrease), which is a binary classification task; (b) predicting the actual price, which is a regression task.  

At this stage of the experiment, 176 companies were grouped into 23 industry sectors. We randomly selected 9 economic sectors and, within each sector, randomly chose two companies. Table~\ref{sectors_tickers_table} lists the selected sectors and companies (tickers) that participated in the computational experiment.  

\begin{figure}[h]
    \centering
    \includegraphics[width=1.0\textwidth]{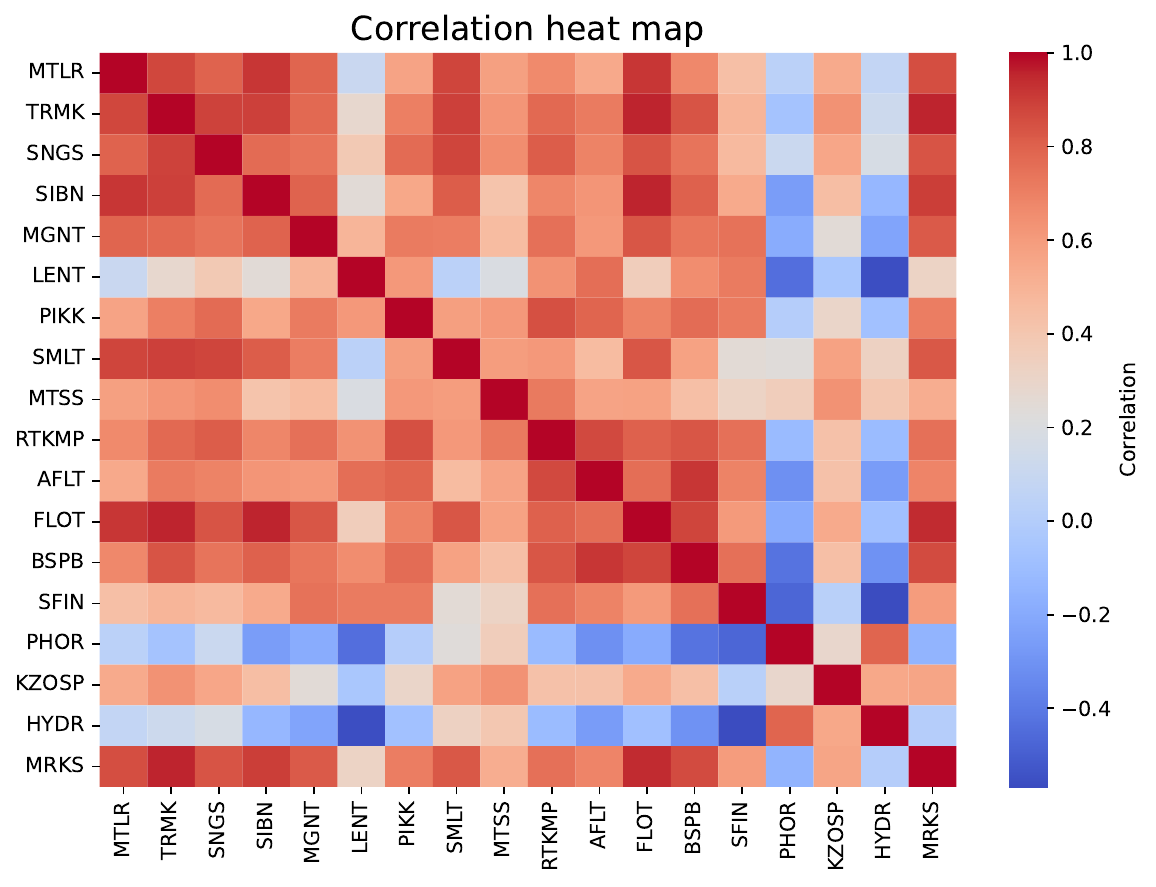} 
    \caption{The correlations heatmap for 18 assets (close price).} 
    \label{fig:corr_map} 
\end{figure}

\begin{figure}[h]
    \centering
    \includegraphics[width=1.0\textwidth]{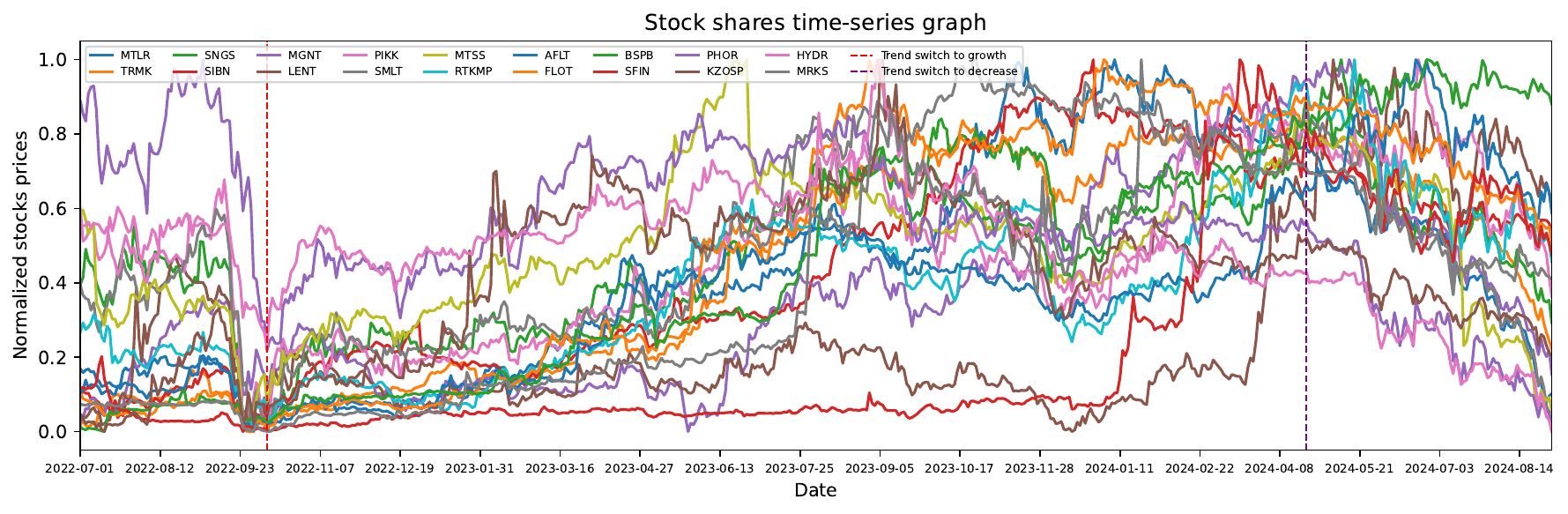} 
    \caption{Normalized close prices of assets. Market phase transition dates denoted by vertical dashed lines.} 
    \label{fig:time_serias} 
\end{figure}

Table~\ref{table:time_series_description} provides statistical data on the closing price time series of the selected assets. The correlation heat map of the closing price time series is shown in Figure~\ref{fig:corr_map}. An interesting feature of the examined period is that the market underwent two phase shifts — from a general price decline to growth and back again — as indicated by the vertical lines in Figure~\ref{fig:time_serias}.

\begin{table}[ht]
\centering
\begin{tabular}{|l|l|}
\hline
Sector & Company (ticker)\\\hline
Metal and Mining      & Mechel (MLTR), TMK-Group (TRMK)    \\ \hline
Oil and Gas & Surgutneftegas (SNGS), Gaspromneft (SIBN)                            \\ \hline
Consumer sector  & Magnit (MGNT), Lenta (LENT)                               \\\hline
Construction & PIK (PIKK), Samolet (SMLT)\\\hline
Telecommunications        & MTS (MTSS), Rostelecom (RTKMP)\\\hline
Transport  & AEROFLOT (AFLT), Sovcomflot (FLOT) \\\hline
Finance & Bank Saint-Petersburg (BSPB), SFI (SFIN) \\\hline
Chemical Industry & Phosagro (PHOR), Kazanorgsintez (KZOSP) \\\hline
Power Engineering  & Rushydro (HYDR), Rosseti Center (MRKC) \\\hline
\end{tabular}
\caption{Economic sectors and companies (tickers) included into the dataset}
\label{sectors_tickers_table}
\end{table}

\begin{table}[ht]
    \centering
    \begin{adjustbox}{width=\textwidth}
        \begin{tabular}{|l|r|r|r|r|r|r|r|}
        \hline
\textbf{Ticker} & \textbf{Mean} & \textbf{Std} & \textbf{Min} & \textbf{Max} & \textbf{Q25} & \textbf{Q50} & \textbf{Q75} \\\hline

MTLR & 191.8245 & 72.5652 & 81.2800 & 332.8800 & 123.8500 & 187.6700 & 251.6400 \\\hline
TRMK & 153.1245 & 64.9362 & 55.8200 & 271.0000 & 87.1400 & 166.4200 & 218.7800 \\\hline
SNGS & 27.0104  & 4.0119  & 17.3500 & 36.9600  & 23.7750 & 27.3300  & 30.0250 \\\hline
SIBN & 601.5097 & 163.9205 & 335.5500 & 934.2500 & 452.0500 & 582.6500 & 748.9000 \\\hline
MGNT & 5691.6429 & 1161.7684 & 4040.0000 & 8444.0000 & 4665.0000 & 5495.0000 & 6375.0000 \\\hline
LENT & 814.3870 & 154.9502 & 650.0000 & 1263.0000 & 716.5000 & 749.0000 & 843.5000 \\\hline
PIKK & 732.6617 & 94.8650 & 518.0000 & 955.5000 & 656.7000 & 732.9000 & 811.5000 \\\hline
SMLT & 3120.8996 & 594.1018 & 1926.5000 & 4145.5000 & 2572.0000 & 3045.0000 & 3713.0000 \\\hline
MTSS & 264.5382 & 32.0791 & 183.0000 & 346.9500 & 239.0000 & 266.2500 & 289.7500 \\\hline
RTKMP & 68.1797 & 9.2753 & 52.2500 & 92.1000 & 60.4500 & 68.0000 & 74.7000 \\\hline
AFLT & 38.1316 & 10.3131 & 22.4400 & 64.4000 & 27.9700 & 38.8800 & 44.1200 \\\hline
FLOT & 88.0111 & 39.5834 & 29.9200 & 149.3000 & 42.1000 & 97.2000 & 124.1800 \\\hline
BSPB & 211.1501 & 101.2533 & 67.5700 & 387.6800 & 100.8400 & 210.9900 & 295.3400 \\\hline
SFIN & 762.9939 & 428.5679 & 425.8000 & 1975.0000 & 497.4000 & 518.0000 & 992.0000 \\\hline
PHOR & 6774.6040 & 618.1977 & 4997.0000 & 8153.0000 & 6416.0000 & 6763.0000 & 7278.0000 \\\hline
KZOSP & 25.8603 & 5.2029 & 15.3500 & 40.5700 & 21.9400 & 27.0700 & 29.8500 \\\hline
HYDR & 0.7697 & 0.0810 & 0.5178 & 1.0278 & 0.7318 & 0.7721 & 0.8210 \\\hline
MRKS & 0.5247 & 0.2382 & 0.2025 & 1.0745 & 0.2735 & 0.5550 & 0.7475 \\\hline
\end{tabular}
\end{adjustbox}
\caption{Descriptive characteristics for company shares}\label{table:time_series_description} 
\end{table}

To evaluate prediction quality in the classification task, we used the \textit{Accuracy} metric, while for regression, we employed \textit{MAPE} (Mean Absolute Percentage Error). The choice of these metrics is justified by the nature of the tasks. In classification, the model must accurately predict the price movement direction either an increase (denoted by "+") or a decrease (denoted by "$-$"). The \textit{MAPE} metric is best suited for assessing regression quality within the financial domain: it represents the average deviation from the asset's actual price in percentage terms, making it easily interpretable in monetary value. 

\begin{figure}[h]
    \centering
    \includegraphics[width=1.0\textwidth]{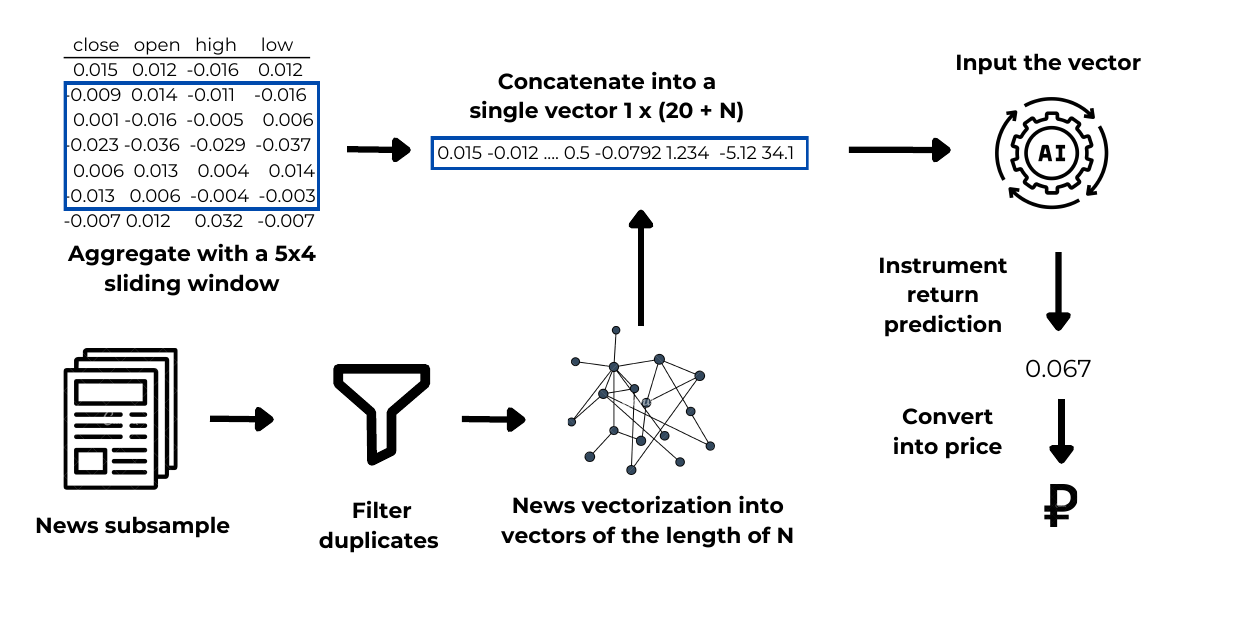} 
    \caption{Pipeline for a single and dual modalities models.} 
    \label{fig:pipeline} 
\end{figure}

Figure~\ref{fig:pipeline} illustrates the model development process for utilizing one and two modalities.  

As the input parameter, the model received a return vector of the asset, calculated based on the closing price (close) over the previous five trading sessions:
\begin{align}\label{eq1}
    Return(d+1) = \frac{close(d+1)}{close(d)} - 1.
\end{align}

The model's output was a prediction for the next trading session.  

To assess the accuracy of predicting the price movement direction, the predicted class was determined by the sign ($\pm$) of the forecasted return value, as the return of an asset represents the relative rate of change. Thus, a positive return indicates a price increase, while a negative return signifies a decline. To evaluate the quality of the asset price forecast, the predicted return vector was converted into price (in Russian rubles):
\begin{equation}\label{eq2}
    price(d+1) = (Return(d+1) + 1) \cdot price(d).
\end{equation}

The pointwise predicted price vector, obtained through transformation, was compared to the historical price vector of assets using the \textit{MAPE} metric.  

The choice of return (rather than price) as the target variable for the predictive model is justified by the fact that when prices exceed historical highs (or fall below historical lows) during market growth (or decline), the applicability of traditional methods becomes limited.  

Based on this reasoning, candlestick characteristics (close, open, high, and low prices) were considered in the form of \textit{relative price changes}, calculated using a formula similar to Eq.~\eqref{eq1}.  

Next, a rolling window of five trading days was applied to the relative price changes to form a vector-row, which was then fed into the predictive model. As a result, the model receives a vector of 20~parameters as input and predicts a single output value — the return of the instrument at the end of the next trading session.

\subsection{The Dual-Modality Approach}

For the experiment involving news flow, we selected news articles relevant to the analyzed assets based on keyword matching (Table~\ref{sectors_tickers_table}). The keywords were chosen as the top 30~words extracted using the TF-IDF method. This method determines the importance of words in a text by considering their frequency of occurrence and uniqueness across the entire corpus. An example of keywords extracted using TF-IDF is presented in Table~\ref{table:key_words_by_ticker}.

\begin{table}[]
    \centering
    \begin{adjustbox}{width=\textwidth}
        \begin{tabular}{|l|l|}
            \hline
            Ticker & Keywords\\\hline
            MTLR & mechel, mining, ore, raw materials, energy, ferroalloys, coal \\\hline
            SNGS & gas, geological exploration, oil, Surgutneftegas, petroleum products, electricity, drilling \\\hline
            SMLT & rent, development, developer, real estate, construction, Moscow region, residential areas\\\hline
        \end{tabular}
    \end{adjustbox}
    \caption{Keywords by companies extracted from their descriptions}\label{table:key_words_by_ticker}
\end{table}

\begin{table}[]
    \centering
    \selectlanguage{russian}
    \begin{adjustbox}{width=\textwidth}
        \begin{tabular}{|l|l|}
            \hline
            Ticker & Keywords \\\hline
            MTLR & мечел, метчел, мечал, mechel, Mchel, ферросплавы, фурросплав \\\hline
            SNGS & сургутнефтегаз, surgutneftegaz, surgut, сурнефтегаз, сургаз, cургут, сур-нфтгз \\\hline
            SMLT & самолет, smlt, samolet, samalet, Самлет\\\hline
        \end{tabular}
    \end{adjustbox}
    \selectlanguage{english}
    \caption{Complementary keywords generated.
    }\label{table:key_words_permuts}
\end{table}

\begin{figure}[h]
    \centering
    \includegraphics[width=1.0\textwidth]{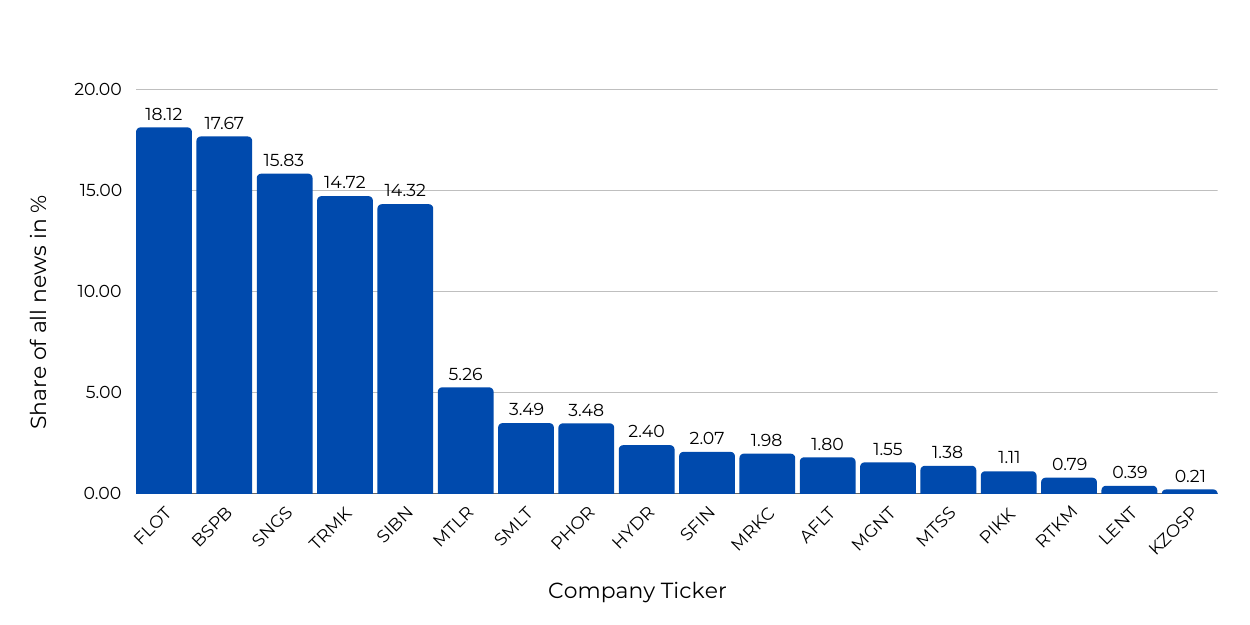}
    \caption{The distribution of news articles by company after filtration (in percents).} 
    \label{fig:news} 
\end{figure}

After obtaining the list of keywords using the TF-IDF method, we further expanded it with the help of the ChatGPT-4o model. This allowed us to increase keyword variability through permutations, letter substitutions, and modifications of word endings (Table~\ref{table:key_words_permuts}).
The selected news articles for each company (ticker) were converted into vectors and filtered to remove duplicates.  
Figure~\ref{fig:news} presents a distribution chart of the news articles for the companies after filtration.
As a vectorizer for the Russian language news stream, we employed two models: RuBERT~\cite{Kuratov(2019)} and Qwen~\cite{nikolichvikhr(2024)}.

While working with the news stream, we encountered two main challenges. The first challenge is the problem of news rewriting, which necessitates filtering out duplicate articles. To ensure that our model accounts for each news article only once, it is essential to implement a duplicate identification algorithm. 
The second challenge is to determinate an asset on which is affected the news article.
This problem can be framed as a classification task, where tickers serve as class labels.

To address the issue of news rewriting, we designed a Siamese neural network. We constructed a training dataset using the GigaChat API as follows: for each article, three paraphrased versions of both the title and body were generated. Then, pairs were randomly formed in equal proportion from the original and paraphrased news articles and their titles. The Siamese neural network was designed as follows: a pair of news articles is fed as input, and vector representations of the articles are extracted using the RuBERT model~\cite{Kuratov(2019)}. The two vectors are then concatenated, and the resulting vector is passed through a fully connected neural network (MLP). To determine the optimal depth of the MLP model, we conducted a series of experiments, evaluating both prediction accuracy and news stream processing time. Based on the results, we selected the MLP architecture with three layers.
The filtered news articles are then converted into vectors so that duplicate classification can be performed in a one-shot mode when new articles arrive. This approach reduces both the processing time of the news stream and the computational resources required (in our case, a GPU V100).

To address the second challenge -- matching news article samples by date and utilizing them for price forecasting -- it is essential to formalize the data selection and prediction process. We assume that the closing price prediction for an asset is made for each trading day at the market opening. In this case, only news articles published before the start of the current trading day are included in the dataset.  

The dataset is formed by grouping news articles based on their publication date. For predicting the price on a given day, only articles published on the previous trading day are used. For example, analytical articles such as those under the "Technical Analysis" section from the "BCS" source, which are published daily before the market opens, are included in the dataset for forecasting the prices of assets analyzed in those reports. This approach ensures that the most relevant information is considered, thereby improving prediction accuracy.

For the dual-modality approach, training sequences were formed by concatenating price return vectors from the previous five days with news stream vectors. The relative price return vectors were constructed similarly to the single-modality experiment, while news articles were selected from the previous trading day based on the chosen asset. These news articles were then transformed into vectors and aggregated.  

If no publications were available on the previous day or before the market opened on the current day, a zero vector was concatenated with the relative price return vector of length 768 for the RuBERT model and 896 for the Vikhr-Qwen2.5-0.5b-Instruct (Qwen) model. Otherwise, the aggregated news vector of the same length was appended. These final vector lengths correspond to the output sizes of the pretrained RuBERT and Qwen models.

In this study, we explored two approaches for aggregating news vectors: vector summation (Sum) and averaged summation (Mean). By vector summation, we mean summing the values of corresponding vector coordinates. In the averaged summation approach, each coordinate of the aggregated vector is assigned the arithmetic mean of the corresponding coordinates across all aggregated vectors. The baseline RuBERT model has a limited context window of 512 tokens. As a result, articles exceeding this limit were either truncated or split for separate processing, meaning that a single news article could correspond to multiple vectors. In contrast, the Qwen model has a significantly larger context window of 32,768 tokens (64 times larger), allowing it to process entire articles without truncation. Next, we compare how different news vectorization methods impact the accuracy of price predictions.

The pointwise predicted return vectors were converted into asset prices using equation~\eqref{eq2}. The prediction quality was evaluated using two metrics: Accuracy and Mean Absolute Percentage Error (MAPE). Accuracy was measured as the proportion of correctly predicted signs of the return vector elements—either positive or negative. The MAPE metric indicates the average percentage deviation of the predicted price from the actual value. This allows us to assess the prediction quality not only in relative terms but also in absolute monetary units (rubles).

\section{Experiment}

In this section, we present the results of computational experiments for two predictive models (single- and dual-modalities). The predictive model was developed using the Transformers framework from the Hugging Face platform. All computations were performed on an NVIDIA V100 GPU.

\subsection{The Single-Modality Approach Performance}

The results of the experiment on predicting return vectors using only time series data for classical and deep learning models are presented in a Table~\ref{table:models_time_series_res}. A Table~\ref{table:avg_models_metrics} provides the averaged prediction quality metrics for all models, sorted in ascending order of the mean absolute percentage error (MAPE) (column "Deviation").  

From the experiment results, it is evident that the recurrent model LSTM achieves the best classification performance (predicting upward or downward trends) and regression accuracy (smallest deviation of the predicted price from the actual price). However, it lags slightly in terms of the mean absolute error metric.

\begin{table}[ht]
    \centering
        \begin{tabular}{|l|r|r|r|}
            \hline
            Model & Accuracy, \% & MAPE, \% \\\hline
            LSTM & \textbf{52.020 }& \textbf{0.397} \\\hline
            XGB & 45.000 & 1.627 \\\hline
            KNN & 46.010 & 1.631 \\\hline
            RF & 48.384 & 1.646 \\\hline
            LinReg & 50.152 & 1.669 \\\hline
            DT & 49.798 & 1.824 \\\hline
        \end{tabular}
    \caption{The Single-Modality approach forecast (time-series) inference metrics: Accuracy and MAPE in percentage.}
    \label{table:avg_models_metrics}
\end{table}

\subsection{The Dual-Modality Approach Performance}

The results of the second experiment, which involved merging the news stream with numerical time series data and comparing the proposed multimodal approach with a forecast based solely on candlestick time series, are presented in the Table~\ref{table:models_multimodal_res}.

The Table~\ref{table:avg_multimodal_metrics} provides the averaged prediction quality metrics for the considered models. The data in this table is sorted by the "Deviation" column in ascending order, reflecting the mean absolute percentage error (MAPE) of the predicted price deviations.

In this second experiment, the LSTM neural network was chosen as the baseline model. We compared different vectorization methods (RuBert, Qwen) and aggregation techniques (Sum, Mean) to evaluate their impact on prediction performance.

\begin{table}[ht]
    \centering
        \begin{tabular}{|l|r|r|r|}
            \hline
            Model & Accuracy, \% & MAPE, \% \\\hline
            LSTM-Qwen-Mean & 48.552 & \textbf{0.256} \\\hline
            LSTM-Qwen-Sum & 46.970 & 0.367 \\\hline
            LSTM & \textbf{52.020 }& 0.397 \\\hline
            LSTM-RuBert-Mean & 49.798 & 0.437 \\\hline
            LSTM-RuBert-Sum & 48.148 & 0.445 \\\hline
        \end{tabular}
    \caption{The Dual-Modality Approach forecast. Accuracy, MAPE in percentage.}
    \label{table:avg_multimodal_metrics}
\end{table}

\begin{figure}[h]
    \centering
    \includegraphics[width=1.0\textwidth]{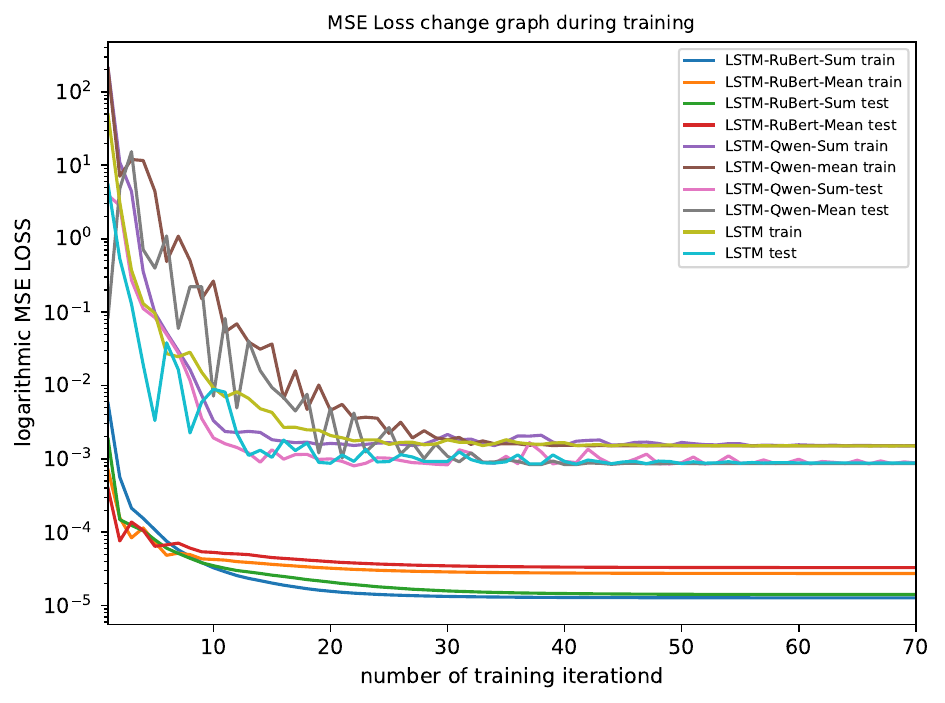}
    \caption{Dependence of the mean squared error function values on the number of training iterations for different models. Training and test sets.} 
    \label{fig:losses_graph} 
\end{figure}

Figure~\ref{fig:losses_graph} shows the dependence of the mean squared error (MSE Loss) function values on the number of training iterations for different models, based on the training set (from July 7, 2022 to March 27, 2024) and the test set (from March 28 to August 30, 2024). The graph indicates that after 30 training epochs, the curves reach a stationary value.

The results from the tables suggest that the forecast based on the vectorized news stream using a large language model outperforms the forecast built solely on candlestick data of assets, demonstrating the smallest deviation of the pointwise price prediction from the actual price vector. Additionally, averaging the vectors (Mean) provides the best results.

The dataset (176 stocks of Russian companies traded on the Moscow Exchange and 79,555 Russian-language financial news articles) collected for the study is available at~\cite{khubiev(2025)}.

\section{Discussion and Conclusion}

As a result of the conducted experiments, we demonstrated that adding a textual modality —analyzing the news stream — positively impacts the accuracy of price prediction. On average, the MAPE metric (the deviation of the predicted price from the actual price) decreases by 55\%: from $0.397$ (LSTM model) to $0.256$ (LSTM-Qwen-Mean model). Additionally, predictions based on vectors obtained using the large language model Vikhr-Qwen2.5-0.5b-Instruct outperformed those based on RuBert. This can be partly attributed to the fact that the Qwen model has a significantly larger context window and is trained on a larger text corpus with support for "Chain-of-Thought" (CoT) reasoning. This enhances the model’s ability to reason and capture complex semantic dependencies within the text. The experimental results indicate that the averaging method (Mean) performed better than summation (Sum) and is the preferred method for aggregating news stream vectors.

At the same time, it is important to note that the test data, on which the final metric values were calculated, covers the period from March 28 to August 30, 2024. During this period, the Russian securities market exhibited a general downward trend. The presence of a clear trend is a significant factor that simplifies the prediction task. However, even in this setting, the proposed multimodal approach proved to be the best among those considered.  

The training and validation of the model for the rewriting task were conducted on news articles whose length did not exceed the context window of the RuBert model. As a result, artifacts related to the context window size only became apparent during the forecasting phase when the news dataset included articles averaging around 290~words in length. For future improvements in news filtering and classification by company, it is necessary to utilize models with a larger context window, such as Qwen.

The collected dataset~\cite{khubiev(2025)} demonstrates good structuring and can be used for fine-tuning large language models in Russian or adapted for the Russian language for applications in the financial sector.


For a quantitative comparison of the proposed model, we conducted a computational experiment based on the approach and metrics from the study~\cite{Kulkova(2023)}. Following the methodology of~\cite{Kulkova(2023)}, we used time series data of stock prices from five major American companies: AAPL, AMZN, GOOGL, NFLX, and TSLA, along with a dataset of English-language news articles labeled by company for the period from October 12, 2012 to January 31, 2020 (Table~\ref{table:comparison}).

\begin{table}[]
    \centering
        \begin{tabular}{|l|c|c|c|c|}
            \hline
            Model & Ticker & R2 & MAPE, \% & MAE \\\hline
            LSTM-Qwen-Mean & AAPL & \textbf{0.989} & \textbf{0.628} & \textbf{0.003} \\\hline
            Baseline & AAPL & 0.947 & 2.333 & 0.018\\\hline
            LSTM-Qwen-Mean & AMZN & \textbf{0.968} & \textbf{1.601} & \textbf{0.013} \\\hline
            Baseline & AMZN & 0.870 & 1.730 & 0.015 \\\hline
            LSTM-Qwen-Mean & GOOGL & \textbf{0.935} & \textbf{1.394} & \textbf{0.008} \\\hline
            Baseline & GOOGL & 0.788 & 2.286 & 0.020 \\\hline
            LSTM-Qwen-Mean & NFLX & \textbf{0.955} & \textbf{2.361} & 0.076 \\\hline
            Baseline & NFLX & 0.919 & 2.512 & \textbf{0.019} \\\hline
            LSTM-Qwen-Mean & TSLA & 0.915 & \textbf{3.206} & \textbf{0.006} \\\hline
            Baseline & TSLA & \textbf{0.930} & 7.423 & 0.034 \\\hline
        \end{tabular}
    \caption{Multimodal approach forecasting metrics in comparison with the approach based on news sentiment score (Baseline) offered by~\cite{Kulkova(2023)}
    }\label{table:comparison}
\end{table}

It is worth noting that the dataset used includes text data in English; therefore, we utilized the original Qwen2.5-0.5b-Instruct model~\cite{qwen2(2023)} for news vectorization. To generate forecasts, we selected and trained the \textit{LSTM-Qwen-Mean} model, as it demonstrated the best overall performance in our study. For evaluation, we used the coefficient of determination (\textit{R2}), mean absolute error (\textit{MAE}), and mean absolute percentage error (\textit{MAPE}). 

Thus, we worked with the same time series and evaluation metrics. Across all metrics, except for \textit{MAE} on NFLX and \textit{R2} on TSLA, the proposed multimodal approach with vector averaging outperformed the best-performing results from the approach in~\cite{Kulkova(2023)}. Based on our computational experiments, we conclude that the proposed multimodal approach demonstrated superior forecasting quality and greater adaptability to both Russian and international markets.

In the future, it is necessary to explore how to incorporate the incoming news stream into the predictive model—specifically, the optimal time window for using news data and the best approach for weighting news messages (e.\,g., adjusting the weight of a news article based on its chronological position in the dataset).

\section*{Acknowledgments}
This work was supported by the grant of the state program of the "Sirius" Federal Territory "Scientific and technological development of the "Sirius" Federal Territory" (Agreement No. 18-03 date 10.09.2024).

\begin{landscape}
\begin{table}
\resizebox{1.5\textheight}{!}{
\begin{tabular}{|l|rr|rr|rr|rr|rr|rr|rr|rr|rr|}
\hline
\multirow{2}{*}{Model} &
\multicolumn{2}{c|}{Metals and Mining} &
\multicolumn{2}{c|}{Oil and Gas} &
\multicolumn{2}{c|}{Consumer Sector} &
\multicolumn{2}{c|}{Construction} &
\multicolumn{2}{c|}{Telecommunications} &
\multicolumn{2}{c|}{Transport} &
\multicolumn{2}{c|}{Finance} &
\multicolumn{2}{c|}{Chemical Industry} &
\multicolumn{2}{c|}{Power Engineering} \\ \cline{2-19} 
 &
  MTLR &
  TRMK &
  SNGS &
  SIBN &
  MGNT &
  LENT &
  PIKK &
  SMLT &
  MTSS &
  RTKMP &
  AFLT &
  FLOT &
  BSPB &
  SFIN &
  PHOR &
  KZOSP &
  HYDR &
  MRKC \\ \hline
\multirow{2}{*}{LSTM} &
  \textbf{56.364} &
  \textbf{56.364} &
  50.303 &
  \textbf{58.182} &
  46.667 &
  \textbf{56.364} &
  49.091 &
  \textbf{53.939} &
  \textbf{56.970} &
  \textbf{55.152} &
  55.152 &
  47.273 &
  46.061 &
  \textbf{49.697} &
  41.818 &
  \textbf{57.576} &
  59.394 &
  40.000 \\
 &
  \textbf{0.410} &
  \textbf{0.362} &
  \textbf{0.352} &
  \textbf{0.341} &
  \textbf{0.331} &
  \textbf{0.371} &
  \textbf{0.484} &
  \textbf{0.328} &
  \textbf{0.541} &
  \textbf{0.246} &
  \textbf{0.419} &
  \textbf{0.258} &
  \textbf{0.410} &
  \textbf{0.447} &
  \textbf{0.231} &
  \textbf{0.458} &
  \textbf{0.380} &
  \textbf{0.768} \\ \hline
\multirow{2}{*}{XGB} &
  40.000 &
  40.909 &
  49.091 &
  40.000 &
  39.091 &
  54.546 &
  40.909 &
  42.727 &
  42.727 &
  45.455 &
  46.364 &
  43.637 &
  49.091 &
  40.000 &
  42.727 &
  49.091 &
  51.182 &
  51.182 \\
 &
  2.089 &
  2.105 &
  1.776 &
  1.766 &
  1.517 &
  2.202 &
  1.565 &
  1.577 &
  1.290 &
  1.299 &
  2.079 &
  2.116 &
  1.612 &
  1.603 &
  1.194 &
  1.198 &
  1.124 &
  1.182 \\ \hline
\multirow{2}{*}{KNN} &
  42.273 &
  38.182 &
  48.182 &
  \textbf{58.182} &
  43.636 &
  39.091 &
  50.909 &
  38.182 &
  40.000 &
  42.723 &
  \textbf{57.273} &
  38.182 &
  50.909 &
  30.909 &
  \textbf{52.723} &
  42.723 &
  \textbf{60.000} &
  49.091 \\
 &
  2.050 &
  2.167 &
  1.775 &
  1.746 &
  1.493 &
  2.178 &
  1.563 &
  1.552 &
  1.306 &
  1.303 &
  2.017 &
  2.124 &
  1.695 &
  1.647 &
  1.149 &
  1.237 &
  1.130 &
  1.225 \\ \hline
\multirow{2}{*}{RF} &
  50.909 &
  47.273 &
  50.000 &
  46.364 &
  49.091 &
  52.723 &
  50.000 &
  46.364 &
  45.455 &
  42.727 &
  52.727 &
  42.727 &
  50.909 &
  39.091 &
  48.182 &
  49.091 &
  48.182 &
  54.545 \\
 &
  2.020 &
  2.154 &
  1.735 &
  1.788 &
  1.519 &
  2.145 &
  1.558 &
  1.539 &
  1.520 &
  1.335 &
  2.062 &
  2.104 &
  1.598 &
  1.743 &
  1.168 &
  1.210 &
  1.214 &
  1.214 \\ \hline
\multirow{2}{*}{LinReg} &
  50.000 &
  49.091 &
  \textbf{60.909} &
  41.818 &
  40.000 &
  51.818 &
  44.545 &
  49.091 &
  53.636 &
  50.909 &
  60.909 &
  45.454 &
  \textbf{54.545} &
  48.182 &
  50.000 &
  46.364 &
  45.455 &
  50.000 \\
 &
  2.029 &
  2.114 &
  1.744 &
  1.839 &
  1.709 &
  2.220 &
  1.637 &
  1.536 &
  1.419 &
  1.355 &
  1.976 &
  2.074 &
  1.602 &
  1.960 &
  1.227 &
  1.217 &
  1.151 &
  1.224 \\ \hline
\multirow{2}{*}{DT} &
  42.727 &
  52.727 &
  52.727 &
  51.818 &
  \textbf{60.000} &
  51.818 &
  \textbf{51.818} &
  41.818 &
  50.000 &
  48.182 &
  51.818 &
  \textbf{49.091} &
  45.455 &
  41.818 &
  45.455 &
  54.545 &
  49.091 &
  \textbf{55.455} \\
 &
  2.679 &
  2.308 &
  1.857 &
  1.813 &
  1.672 &
  2.589 &
  1.592 &
  1.683 &
  1.395 &
  1.411 &
  2.194 &
  2.294 &
  1.829 &
  1.959 &
  1.218 &
  1.581 &
  1.355 &
  1.403 \\ \hline
\end{tabular}%
}
    \caption{Returns vector forecasting metrics with only time-series in use. Accuracy (the upper row), MAPE (the lower row) in percentage}
    \label{table:models_time_series_res}
\end{table}
\end{landscape}

\begin{landscape}
\begin{table}
    \centering
    \caption{The Dual-Modality returns vector forecasting metrics. Accuracy (the upper row), MAPE (the lower row) in percentage.}\label{table:models_multimodal_res}
    \resizebox{1.5\textheight}{!}{
    \begin{tabular}{|l|rr|rr|rr|rr|rr|rr|rr|rr|rr|}
        \hline
        \multirow{2}{*}{Model} & 
        \multicolumn{2}{c|}{Metals and Mining} &
        \multicolumn{2}{c|}{Oil and Gas} &
        \multicolumn{2}{c|}{Consumer Sector} &
        \multicolumn{2}{c|}{Construction} &
        \multicolumn{2}{c|}{Telecommunications} &
        \multicolumn{2}{c|}{Transport} &
        \multicolumn{2}{c|}{Finance} &
        \multicolumn{2}{c|}{Chemical Industry} &
        \multicolumn{2}{c|}{Power Engineering} \\
        \cline{2-19}
        & MTLR & TRMK & SNGS & SIBN & MGNT & LENT & PIKK & SMLT & MTSS & RTKMP & AFLT & FLOT & BSPB & SFIN & PHOR & KZOSP & HYDR & MRKC\\
        \hline
        \multirow{2}{*}{vanilla LSTM}
        & \textbf{56.364\%} & \textbf{56.364\%} 
        & 50.303\% & \textbf{58.182\%} 
        & 46.667\% & \textbf{56.364\%} 
        & 49.091\% & \textbf{53.939\%} 
        & \textbf{56.970\%} & \textbf{55.152\%} 
        & \textbf{55.152\%} & 47.273\% 
        & 46.061\% & 49.697\% 
        & 41.818\% & \textbf{57.576\%} 
        & 59.394\% & 40.000\% 
        \\
        & 0.410\% & 0.362\% 
        & 0.352\% & 0.341\% 
        & 0.331\% & 0.371\% 
        & 0.484\% & 0.328\% 
        & 0.541\% & 0.246\% 
        & 0.419\% & \textbf{0.258\%} 
        & 0.410\% & 0.447\% 
        & 0.231\% & 0.458\% 
        & 0.380\% & 0.768\% 
        \\\hline
        \multirow{2}{*}{LSTM\_RuBert\_SUM}
        & 39.394\% & 35.152\% 
        & 53.939\% & \textbf{58.182\%} 
        & \textbf{53.333\%} & 49.091\% 
        & 50.303\% & 38.788\% 
        & 53.939\% & 49.697\% 
        & 51.515\% & 43.636\% 
        & 47.879\% & 44.848\% 
        & 53.333\% & 42.424\% 
        & 58.788\% & 42.424\% 
        \\
        & 0.409\% & 0.392\% 
        & 0.865\% & 0.265\% 
        & 0.417\% & 0.400\% 
        & 0.462\% & \textbf{0.200\%} 
        & 0.473\% & 0.274\% 
        & 0.641\% & 0.532\% 
        & 0.406\% & 0.445\% 
        & 0.264\% & 0.492\% 
        & 0.326\% & 0.742\% 
        \\\hline
        \multirow{2}{*}{LSTM\_RuBert\_MEAN}
        & 38.788\% & 42.424\% 
        & \textbf{58.182\%} & \textbf{58.182\%} 
        & 47.879\% & 50.909\% 
        & \textbf{57.576\%} & 46.061\% 
        & 55.152\% & 45.455\% 
        & 50.303\% & \textbf{52.121\%} 
        & 50.909\% & 47.273\% 
        & 55.152\% & 41.212\% 
        & 55.758\% & \textbf{43.030\%} 
        \\
        & 0.410\% & \textbf{0.192\%} 
        & 1.824\% & 0.216\% 
        & 0.299\% & 0.359\% 
        & 0.436\% & 0.270\% 
        & 0.368\% & 0.271\% 
        & 0.348\% & 0.262\% 
        & 0.326\% & 0.390\% 
        & 0.238\% & 0.491\% 
        & 0.321\% & 0.839\% 
        \\\hline
        \multirow{2}{*}{LSTM\_QWEN\_SUM}
        & 45.455\% & 36.364\% 
        & 44.848\% & 39.394\% 
        & 46.061\% & 53.333\% 
        & 47.273\% & 36.364\% 
        & 47.879\% & 44.848\% 
        & 45.455\% & 43.636\% 
        & 47.879\% & \textbf{56.970\%} 
        & \textbf{60.000\%} & 48.485\% 
        & 47.879\% & 42.424\% 
        \\
        & 0.522\% & 0.504\% 
        & 0.307\% & 0.368\% 
        & 0.307\% & 0.346\% 
        & 0.529\% & 0.311\% 
        & 0.316\% & \textbf{0.171\%} 
        & 0.259\% & 0.392\% 
        & 0.369\% & \textbf{0.195\%} 
        & 0.354\% & 0.369\% 
        & 0.292\% & 0.660\% 
        \\\hline
        \multirow{2}{*}{LSTM\_QWEN\_MEAN}
        & 52.121\% & 35.758\% 
        & 49.697\% & 47.879\% 
        & 48.485\% & 52.121\% 
        & 53.333\% & 43.030\% 
        & 45.455\% & 44.242\% 
        & 52.121\% & 43.636\% 
        & \textbf{52.121\%} & \textbf{56.970\%} 
        & 44.848\% & 49.697\% 
        & \textbf{61.212\%} & 41.818\% 
        \\
        & \textbf{0.246}\% & 0.419\% 
        & \textbf{0.106\%} & \textbf{0.165\%} 
        & \textbf{0.235\%} & \textbf{0.331\%} 
        & \textbf{0.322\%} & 0.241\% 
        & \textbf{0.193\%} & 0.178\% 
        & \textbf{0.182\%} & 0.345\% 
        & \textbf{0.227\%} & 0.272\% 
        & \textbf{0.219\%} & \textbf{0.352\%} 
        & \textbf{0.178\%} & \textbf{0.543\%} 
        \\\hline
    \end{tabular}
}
\end{table}
\end{landscape}

\bibliographystyle{unsrt}

\end{document}